\documentclass{article}
\usepackage{spconf,amsmath,graphicx}

\usepackage[T1]{fontenc}
\usepackage[utf8]{inputenc}
\usepackage{microtype}

\usepackage{booktabs}
\usepackage{multirow}
\usepackage{latexsym}
\usepackage{stfloats}
\usepackage{algpseudocode}
\newcounter{algoctr}

\usepackage{url}
\usepackage{tikz}
\usetikzlibrary{positioning,arrows.meta,calc,fit,backgrounds}
\usepackage{etoolbox}
\apptocmd{\thebibliography}{\setlength{\itemsep}{0pt}\setlength{\parsep}{0pt}}{}{}
\usepackage{hyperref}   

\newcommand{\parhead}[1]{\vspace{1.5pt}\noindent\textbf{#1}\ }
\title{SheetMind: Actions Set Accuracy, Agents Set the Failure Mode}
\name{\begin{tabular}{c}
 Lyuhao Chen$^{1\ast}$, Xi Cheng$^{2\ast}$, Yanming Kang$^{3}$, Ruiyan Zhu$^{2}$, Ke Liu$^{4}$, Rakesh Chowdary Machineni$^{5}$, \\
 Yulang Fei$^{3}$, Brian Zhu$^{2}$, Daniel Jin$^{2}$, Binze Cai$^{7}$, Zheng Qi$^{6}$, Neeraj Parihar$^{2}$, \\
 Zhoutian Xu$^{2}$, Oliver Gao$^{2}$
\end{tabular}}
\address{$^{1}$Carnegie Mellon University \quad
         $^{2}$Cornell University \quad
         $^{3}$University of Waterloo \quad
         $^{4}$UC Berkeley \\
         $^{5}$University of Michigan, Ann Arbor \quad
         $^{6}$HKUST (Guangzhou) \quad
         $^{7}$Georgia Institute of Technology}
\begin{document}
\maketitle
{\renewcommand{\thefootnote}{\fnsymbol{footnote}}\setcounter{footnote}{1}\footnotetext{Equal contribution.}}
\begin{abstract}
Spreadsheet agents are converging on elaborate multi-agent designs, yet it is unclear how much of their performance comes from the agents rather than from the action interface they share. We answer this with \textit{SheetMind}, a Manager--Action--Reflection framework, in a controlled study over all 221 tasks of the SheetCopilot Benchmark: five architectural variants, four backbones, exact McNemar tests on paired outcomes, and a checker reproducing the official chart and pivot comparisons. Replacing the high-level action API with primitive cell operations costs \textbf{47.1} points ($p<10^{-4}$) and leaves the agent \emph{below} a do-nothing baseline, whereas both extra agents together are worth \textbf{3.2} points: the Reflection Agent adds $+4.5$ ($p{=}0.013$), the Manager $+1.4$ ($p{=}0.68$). Decomposition instead changes \emph{how} the system fails, cutting silently wrong outputs from 33\% to 25\% of tasks ($p{=}0.010$). Capability saturates: GPT-5 and the five-times-cheaper GPT-5-mini are not significantly different (61.1\% vs.\ 58.4\%, $p{=}0.15$), while GPT-3.5 loses 16.3 points and fails differently. A reflector must judge \emph{the step it just took}, not the subtask. SheetMind reaches 61.1\% Pass@1 with GPT-5 on the full SCB-221, against a do-nothing baseline of 9.0\%. Accuracy comes from the operations an agent can name; the agents decide how it fails.
\end{abstract}
\begin{figure}[!t]
    \centering
    \includegraphics[width=0.84\linewidth]{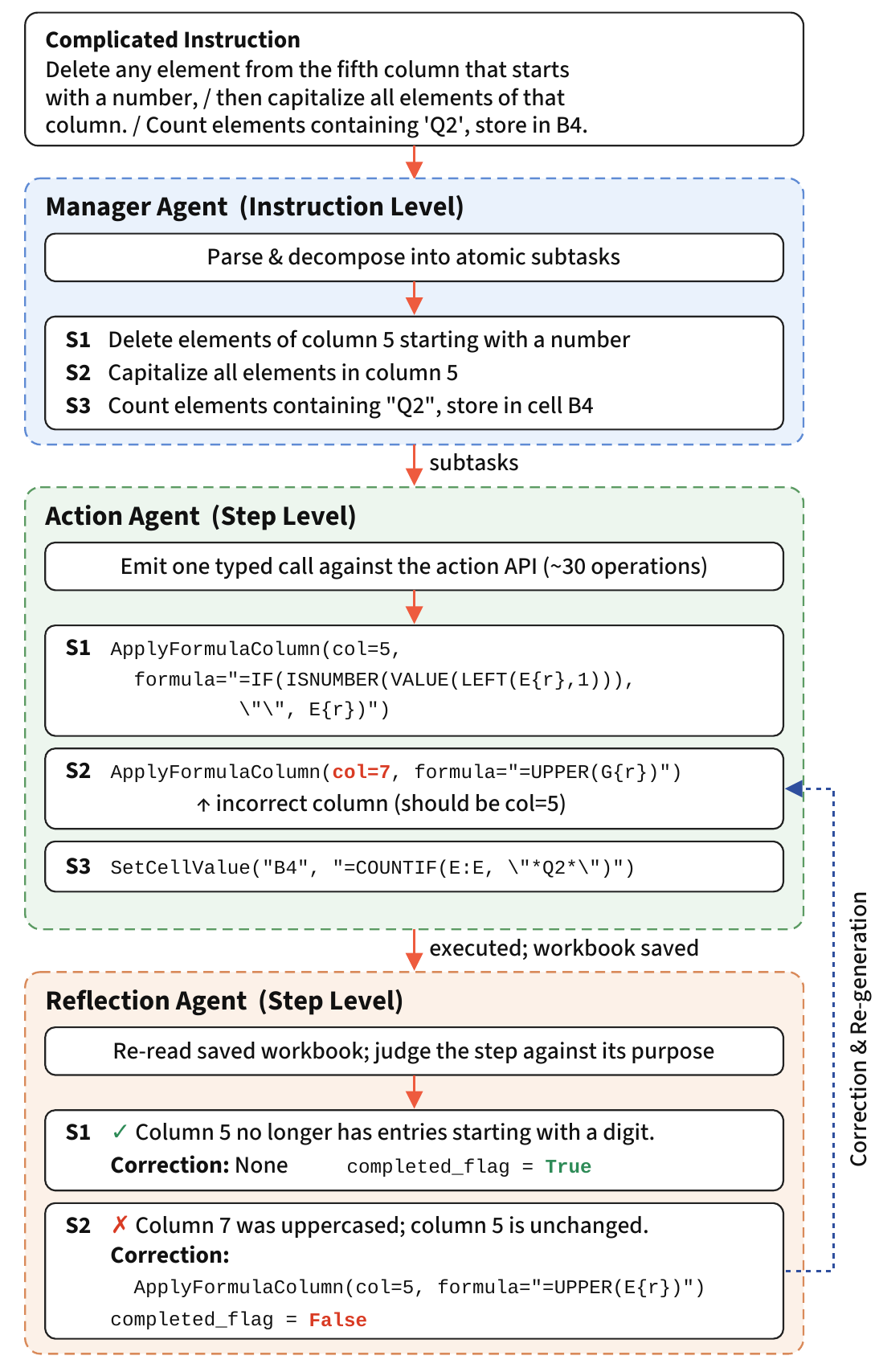}
    \caption{The evaluated loop. The Manager splits the instruction and one iteration budget into subtasks; the Action Agent emits one typed call per step; the workbook on disk, re-read after every execution, is the only observation; the Reflection Agent judges \emph{that step} against its stated purpose (A accept, B/C regenerate). The ablations remove the Manager, remove the Reflection Agent, or replace the API with the eight primitive cell operations.}
    \label{fig:pipeline}
\end{figure}

\section{Introduction}
\label{sec:intro}
Spreadsheets are ubiquitous, yet using them well requires formulas, macros and structured syntax, a barrier for non-technical users. Natural language interfaces lower it, and large language models have made them practical for spreadsheet work~\cite{li2023sheetcopilot}.

The dominant response has been architectural. SheetCopilot~\cite{li2023sheetcopilot} pairs atomic spreadsheet actions with a state-machine planner; SheetBrain~\cite{wang2026sheetbrain} an understand--execute--validate pipeline. Each reports gains, and each varies roles, prompts and the operation set at once. What none isolates is how much of the gain belongs to the \emph{agents} rather than to the \emph{action interface} they share, even though in adjacent domains interface refinement alone lifts a web agent by 26.6 points~\cite{yang2025agentoccam} and the action representation alone lifts 17 LLMs by up to 20~\cite{wang2024codeact,yang2024sweagent}.

We measure both in one experiment: five variants of SheetMind and four backbones over all 221 tasks of the SheetCopilot Benchmark (SCB-221), with exact McNemar tests~\cite{dietterich1998tests} on paired per-task outcomes. We find that removing both extra agents costs 3.2 points, while replacing the high-level action API with the primitive one costs 47.1 and drops the system \emph{below} a baseline that makes no edit at all. Extra agents buy little Pass@1, only the Reflection Agent is significant, but decomposition changes \emph{how} the system fails, replacing silently wrong outputs by self-reported incompletion. Moving from GPT-5-mini to GPT-5 buys six tasks out of 221 ($p{=}0.15$).

Contributions. (1)~\textbf{Action abstraction dominates architecture} by an order of magnitude, and the primitive action space has \emph{negative} net value against a do-nothing reference (Sec.~\ref{sec:action}). (2)~\textbf{Reflection raises Pass@1, decomposition changes the failure mode}: $+4.5$ ($p{=}0.013$) against $+1.4$ ($p{=}0.68$), but the Manager replaces silently wrong outputs by self-reported incomplete ones (33\%$\to$25\%, $p{=}0.010$), helping on formula and entry tasks and hurting on pivot tables; a reflector must judge the step, not the subtask (Sec.~\ref{sec:agents}). (3)~\textbf{Capability saturates}: GPT-5 and GPT-5-mini are not significantly different and share 83 of $\sim$90 failures, while GPT-3.5 fails differently rather than merely more often (Sec.~\ref{sec:backbone}).

\section{Related Work}
\label{sec:related}
\parhead{LLM agents for spreadsheets.}
SheetCopilot~\cite{li2023sheetcopilot} introduced SCB-221 with an atomic action set and a state-machine planner. Recent work has migrated to harder benchmarks~\cite{ma2024spreadsheetbench,zhu2026spreadsheetbench2,wang2026sheetbrain}, but we retain SCB-221 deliberately: it is the only spreadsheet benchmark with a per-task reference workbook and an official checker, which is what makes 221 \emph{paired} outcomes, and hence our significance tests, possible. All of these systems vary architecture and action set together; we vary one at a time.

\parhead{Action space versus architecture.}
That an agent is bounded by its interface is implicit in tool-use work~\cite{schick2023toolformer,patil2024gorilla}, and recent studies isolate it: AgentOccam~\cite{yang2025agentoccam} obtains $+26.6$ points on web tasks from observation/action refinement alone; CodeAct~\cite{wang2024codeact} up to $+20$ by making actions executable code; SWE-agent~\cite{yang2024sweagent} argues that the agent--computer interface is the enabling component. No comparable isolation exists for spreadsheets.

\parhead{Reflection and decomposition.}
Self-critique~\cite{shinn2023reflexion,madaan2023selfrefine} is standard; Mobile-Agent~\cite{wang2024mobileagent} contributes the post-execution A/B/C verdict, and Mobile-Agent-v2~\cite{wang2024mobileagentv2} reports large gains from the planner/decision/reflection triple. The critical literature is equally established: LLMs do not reliably self-correct without an external signal~\cite{huang2024selfcorrect}, and the bottleneck is error \emph{localization} rather than correction~\cite{tyen2024errorlocation}. Decomposition exists to bring each step within what the executor can do~\cite{zhou2023leasttomost,khot2023decomposed}; ADaPT~\cite{prasad2024adapt} applies it only when the executor cannot carry a step, and compound system analyses show that adding LLM calls is non-monotone~\cite{chen2024morecalls,cemri2025masfail}.

\section{System and Ablated Variants}
\label{sec:method}
SheetMind coordinates three LLM agents over a workbook that is re-read from disk every turn (Fig.~\ref{fig:pipeline}); Algorithm~\ref{alg:loop} gives the loop.

\begin{figure}[t]
\small
\hrule height 0.8pt
\vspace{3pt}
\noindent\refstepcounter{algoctr}\textbf{Algorithm \thealgoctr}\; \textbf{The evaluated SheetMind loop}\label{alg:loop}
\vspace{2pt}
\hrule height 0.4pt
\vspace{2pt}
\noindent\textbf{Input:} instruction $I$, workbook $W_0$ on disk, iteration budget $B$\\
\textbf{Output:} the workbook on disk
\vspace{2pt}
\hrule height 0.4pt
\vspace{2pt}
\begin{algorithmic}[1]
\State $T \gets \textsc{Manager}(I)$ \Comment{without the Manager: $T \gets \{I\}$}
\State split $B$ evenly over $T$; unused budget rolls over
\For{each subtask $t_i \in T$ with share $b_i$}
    \State $H \gets \emptyset$;\; $\mathit{rejects} \gets 0$
    \While{$b_i > 0$ \textbf{and} $t_i$ not done}
        \State $W \gets \textsc{Read}(\text{disk})$;\; $b_i \gets b_i - 1$
        \State $a \gets \textsc{ActionAgent}(t_i, W, H)$ \Comment{one typed API call}
        \State $W' \gets \textsc{Execute}(a, W)$;\; \textsc{Save}$(W')$;\; $H \gets H \cup \{a\}$
        \State $v \gets \textsc{Reflection}(a.\mathit{purpose}, W, W')$ \Comment{A / B / C}
        \If{$v = \texttt{A}$ \textbf{or} $\mathit{rejects} = 2$}
            \State $\mathit{rejects} \gets 0$;\; done $\gets a.\mathit{completed}$
        \Else
            \State $H \gets H \cup \{v\}$;\; $\mathit{rejects} \gets \mathit{rejects} + 1$
        \EndIf
    \EndWhile
\EndFor
\State \Return the workbook on disk
\end{algorithmic}
\vspace{2pt}
\hrule height 0.2pt
\end{figure}

\parhead{Action Agent and its action space.}
The Action Agent emits exactly one typed call per turn against a documented API of $\sim$30 high-level operations (\texttt{Apply\-Formula\-Column}, \texttt{Create\-Summary\-Table}, \texttt{Create\-Chart}, \texttt{Set\-Con\-di\-tion\-al\-For\-mat}, \texttt{Date\-Extract}, \texttt{Filter}, \texttt{Sort}, sheet management), each compiling to many primitive cell writes. A parser rejects malformed calls and requests a regeneration, so an accepted action is executable; semantic correctness is left to reflection. Formulas are evaluated by a spreadsheet engine at write time and materialised as values. The ablated \textbf{base} action space substitutes the eight primitive operations of the original benchmark agent (select, drag, set, read, format, \ldots) together with a prompt documenting only those; nothing else changes.

\parhead{Manager Agent.}
The Manager maps an instruction $I$ to an ordered subtask list $T=\{t_1,\dots,t_k\}$, each intended to be one atomic spreadsheet operation. A single iteration budget $B$ is divided evenly across subtasks with rollover: subtask $i$ of $k$ receives
\begin{equation}
b_i=\max\!\big(2,\;\lfloor B_{\text{rem}}/(k-i+1)\rfloor\big)
\label{eq:budget}
\end{equation}
of the budget $B_{\text{rem}}$ left by its predecessors. This equalises the LLM-call budget across variants.

\parhead{Reflection Agent.}
After each execution the Reflection Agent compares the pre- and post-execution workbook against \emph{the executed action's own stated purpose} and returns \texttt{A} (proceed), \texttt{B} (incorrect) or \texttt{C} (no change); on \texttt{B}/\texttt{C} the Action Agent regenerates with the verdict appended, and an action rejected twice in a row is accepted so the loop advances. Two properties matter (Sec.~\ref{sec:agents}): the verdict concerns one step, not the subtask, and a rejected step stays in the history because it has already been executed and saved.

\section{Experimental Setup}
\label{sec:setup}
\parhead{Benchmark and metric.}
SCB-221~\cite{li2023sheetcopilot} contains 221 tasks in six categories, annotated with a reference decomposition length (109 single-step, 112 multi-step). We report Pass@1 from an openpyxl port of the official Excel-driven checker that keeps its chart comparison (type, title, legend, axis titles, series by dereferenced values) and checks pivot tables by rendered content. A failed task is \emph{detectable} if a subtask did not terminate or a required artifact is absent, and \emph{silent-wrong} otherwise (Sec.~\ref{sec:agents}). We deliberately do \textbf{not} report Exec@1: our harness copies the input workbook to the output path before the agent runs and catches per-action exceptions, so ``the file exists and opens'' holds by construction, which is a property of the harness rather than of the system.

\parhead{Variants and backbones.}
We evaluate five variants, \textbf{A} (Action only), \textbf{A+R}, \textbf{M+A}, \textbf{M+A+R} (full), and \textbf{A-base} (Action only, primitive action space), on four backbones: GPT-3.5-Turbo, GPT-5-mini, GPT-5 and, for the full system only, the open-weight Llama-3.1-70B-Instruct. All runs cover all 221 tasks under the same iteration budget and a 1200\,s per-task guard; a \textbf{no-op} reference scores the unmodified input workbooks.

\parhead{Statistics.}
Variants are paired on tasks, so we report exact McNemar tests~\cite{dietterich1998tests} over discordant pairs: with $b$ tasks passed only by one variant and $c$ only by the other, the exact test is the two-sided binomial on the $b{+}c$ discordant pairs. At $n{=}221$ and the observed discordance, effects of $\geq$5 points are detectable; our null results ($\pm$1--3 points) should be read as ``no effect of practically relevant size'' rather than proof of equality~\cite{card2020littlepower}.

\section{Results}
\label{sec:results}

\begin{figure}[t]
\centering
\includegraphics[width=0.9\linewidth]{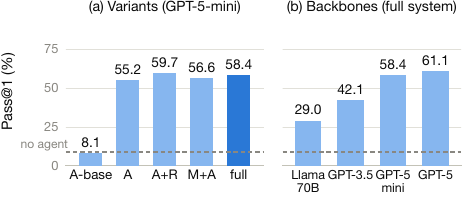}
\caption{Pass@1 over all 221 tasks, (a)~per variant with GPT-5-mini and (b)~for the full system (M$+$A$+$R) per backbone; the dashed line is the no-agent reference (9.0\%).}
\label{fig:pass}
\end{figure}

\begin{figure}[t]
\centering
\includegraphics[width=\linewidth]{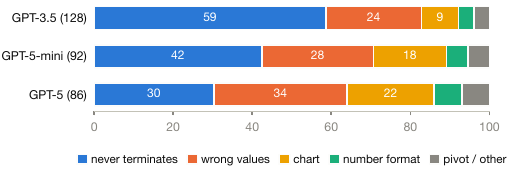}
\caption{Failure composition of the full system per GPT backbone (count in brackets).}
\label{fig:fail}
\end{figure}

Table~\ref{tab:main} places SheetMind against published SCB-221 results. With GPT-5 it reaches 61.1\% Pass@1; with GPT-3.5 it scores 42.1\%, 2.2 points below SheetCopilot on the same backbone. Since execution backends differ across systems, the table is placement rather than ranking. We next break this result down by action space, agents and backbone (Figs.~\ref{fig:pass}--\ref{fig:fail}, Table~\ref{tab:mcnemar}).

\begin{table}[t]
\centering
\small
\begin{tabular}{llc}
\toprule
\textbf{Backbone} & \textbf{Method} & \textbf{Pass@1}$\uparrow$ \\
\midrule
--  & No agent (unmodified input) & 9.0 \\
\midrule
\multirow{3}{*}{GPT-3.5}
     & VBA$^\dagger$            & 16.3 \\
     & SheetCopilot$^\dagger$   & \textbf{44.3} \\
     & SheetMind (ours)         & \underline{42.1} \\
\midrule
\multirow{3}{*}{Llama-3.1-70B}
     & SheetCopilot$^\ddagger$  & 21.5 \\
     & MaxMind$^\ddagger$       & \underline{23.5} \\
     & SheetMind (ours)         & \textbf{29.0} \\
\midrule
\multirow{3}{*}{GPT-4 class}
     & SheetCopilot$^{\dagger *}$ & \underline{55.0} \\
     & OS-Copilot$^{\S *}$        & \textbf{60.0} \\
     & SheetMind (ours, GPT-4.1)$^{\P}$ & \textbf{60.0} \\
\midrule
GPT-5 & SheetMind (ours)        & \textbf{61.1} \\
\bottomrule
\end{tabular}
\caption{Pass@1 (\%) on SCB-221. Best per backbone in bold, second best underlined. $^\dagger$as reported by SheetCopilot~\cite{li2023sheetcopilot}; $^\ddagger$as reported by MaxMind~\cite{dong2024maxmind}; $^\S$as reported by OS-Copilot~\cite{wu2024oscopilot}; $^*$GPT-4 on the 20-task subset of SCB-221; $^\P$GPT-4.1 on 20 tasks of SCB-221 (Sec.~\ref{sec:backbone}). SheetMind rows use the openpyxl port of the official checker (Sec.~\ref{sec:setup}).}
\label{tab:main}
\end{table}

\begin{table}[t]
\centering
\small
\setlength{\tabcolsep}{2.6pt}
\begin{tabular}{lcccccc}
\toprule
 & $\Delta$\textbf{All} & $\Delta$\textbf{Single} & $\Delta$\textbf{Multi} & $p$ & \textbf{Silent} & \textbf{Det.} \\
\midrule
\multicolumn{7}{l}{\emph{Variants (GPT-5-mini), $\Delta$ vs.\ A (Pass@1 55.2)}} \\
A-base  & $-47.1$ & $-59.6$ & $-34.8$ & $<10^{-4}$ & --   & --   \\
A       & --      & --      & --      & --         & 33.0 & 11.8 \\
A+R     & $+4.5$  & $+4.6$  & $+4.5$  & $0.013$    & 35.3 & 5.0  \\
M+A     & $+1.4$  & $+1.8$  & $+0.9$  & $0.678$    & 24.9 & 18.6 \\
M+A+R   & $+3.2$  & $+4.6$  & $+1.8$  & $0.118$    & 23.5 & 18.1 \\
\midrule
\multicolumn{7}{l}{\emph{Backbones (full system), $\Delta$ vs.\ GPT-5-mini (58.4)}} \\
GPT-3.5 & $-16.3$ & $-19.3$ & $-13.4$ & $<10^{-4}$ & -- & -- \\
GPT-5   & $+2.7$  & $+2.8$  & $+2.7$  & $0.146$    & -- & -- \\
\bottomrule
\end{tabular}
\caption{Paired Pass@1 difference (points) to the reference row over all 221 tasks and over the 109 single-step / 112 multi-step tasks of SheetCopilot's annotation; $p$ from exact McNemar tests on all tasks. Silent/Det.: share of the 221 tasks that are \emph{silent-wrong} (completion reported, content wrong) or \emph{detectable} (a subtask did not terminate or an artifact was never created); the Manager moves tasks from silent to detectable ($p{=}0.010$; $p<10^{-4}$ on top of reflection).}
\label{tab:mcnemar}
\end{table}

\subsection{The action space dominates}
\label{sec:action}
Substituting the primitive action space for the high-level one costs 47.1 points ($p<10^{-4}$), against 3.2 points for the two additional agents and 4.5 for the better of them, an order of magnitude in the same experiment.

The primitive-action variant passes 18 tasks; \emph{all 18} are tasks that the unmodified input workbook already passes, and it breaks two of them. Its net contribution over editing nothing is therefore $-2$ tasks: with primitive operations the agent does not solve a single task on its own. The effect is categorical: the same LLM that solves nothing with primitives passes 55.2\% with a high-level API.

\subsection{Reflection raises Pass@1; decomposition changes the failure mode}
\label{sec:agents}
The Reflection Agent is worth $+4.5$ points ($p{=}0.013$), the Manager $+1.4$ ($p{=}0.678$), and the two together $+3.2$ ($p{=}0.118$), short of the sum of their separate effects. Both attack the same failure, an instruction longer than one step, from opposite sides, so they overlap, as ADaPT's ``decompose only when the executor cannot proceed''~\cite{prasad2024adapt} and the non-monotone scaling of compound systems~\cite{chen2024morecalls} predict.

\parhead{Where decomposition really helps.}
Split failures into \emph{silent-wrong} and \emph{detectable} (Table~\ref{tab:mcnemar}; 39 non-terminating tasks plus one never-created chart for the full system): the first hands the user a plausible workbook whose content is wrong, the second the system can report. Silent-wrong outputs fall by 8.1 points with the Manager (Table~\ref{tab:mcnemar}; 31 removed, 13 added; $p{=}0.010$) and by 11.8 on top of reflection (31 vs.\ 5; $p<10^{-4}$); most of the removed ones become detectable failures, not passes. The schedule itself is not the limit: re-running the full system on Llama-3.1-70B with an on-demand pool instead of the even split moves three tasks (29.0\% to 30.3\%, 14 wins/11 losses, $p{=}0.69$). A+R has the highest Pass@1 but returns a plausible wrong file on 35\% of tasks, M+A+R on 24\%, which matters for spreadsheets, whose costliest error is a silent wrong number. The gain sits where an instruction is a sequence of cell operations (Formula $+6.2$ points, 9 wins/3 losses, $n{=}97$; Entry \& manipulation $+3.0$, 13/8, $n{=}165$) and is negative where the API already does the whole task in one call: Pivot Table $-7.7$ (0/3, $n{=}39$), Charts $0.0$. This is ADaPT's condition~\cite{prasad2024adapt}: decomposition pays only for what the executor cannot do in one step, and a high-level API makes a pivot table or chart one step.

\parhead{The reflector's granularity decides its sign.}
Two formulations of the same Reflection Agent differ only in the question asked. A \emph{subtask-scoped} reflector, asked whether the workbook now satisfies the subtask, marks every correct intermediate step as a failure because later steps are still undone: over 571 verdicts, 76\% are negative and \textbf{95\% of the rejections follow an action that executed successfully}. Dropping the rejected step from the history, the GUI-agent convention~\cite{wang2024mobileagent}, compounds this once the action is saved: the feedback then cites a stale action in 100\% of 384 messages. The cost is that 32\% of actions repeat the previous one and 39\% of subtasks never terminate, against 12\% and 18\% with no reflector. A \emph{step-scoped} reflector, asked only whether the executed action achieved its own stated purpose, retaining rejected steps, and accepting any action rejected twice, yields 4\% negative verdicts and 2\% repeated actions, and is the configuration measured throughout. Reflection contributes error \emph{localization}~\cite{tyen2024errorlocation}; a question about the goal rather than the step localizes nothing.

\subsection{Capability saturates above GPT-3.5 and reshapes the failures}
\label{sec:backbone}
GPT-5 and GPT-5-mini differ by six tasks out of 221 ($+2.7$ points for GPT-5, $p{=}0.146$) and fail on 83 of the same tasks; 81 tasks are failed by all three backbones. GPT-3.5 is 16.3 points behind ($p<10^{-4}$) and fails differently: 59\% of its failures never terminate and at least 69 of them repeat one action more than twice. The open-weight Llama-3.1-70B reaches 29.0\% with the same system, below GPT-3.5 but above both published Llama results on the benchmark (Table~\ref{tab:main}), and, like GPT-3.5, it reports completion on only 16\% of tasks. Above the threshold the model is not the binding constraint, consistent with agentic competence being gated by a threshold rather than scaling smoothly~\cite{liu2024agentbench}. GPT-4.1 on 20 randomly drawn tasks passes 12 of them; on the same tasks GPT-3.5 passes 9, and GPT-5-mini and GPT-5 pass 13 each, so the plateau already begins with the GPT-4 generation.

\parhead{Failure composition shifts with capability.}
Of GPT-5-mini's 92 failures, 39 never terminate, 26 return wrong values, 17 a wrong or missing chart, 5 a wrong number format, 4 a wrong pivot table and 1 other. Across backbones the non-terminating share falls with capability (59\%, 42\%, 30\%) while the wrong-value share rises (24\%, 28\%, 34\%): the stronger the model, the more of what remains is a silent content error rather than a visible stall, exactly the failure mode the Manager converts (Sec.~\ref{sec:agents}).

\section{Conclusion}
\label{sec:conclusion}
On the full SheetCopilot Benchmark, under the official chart and pivot comparisons, the action space's abstraction level is worth 47.1 points and two additional agents 3.2; only the Reflection Agent raises Pass@1; the Manager instead turns silently wrong outputs into self-reported incomplete ones and pays only above the API; and above GPT-3.5 the backbone stops mattering. SheetMind reaches 61.1\% Pass@1 with GPT-5 (58.4\% with GPT-5-mini). Accuracy comes from the operations an agent can name; the agents decide how it fails.

\vfill\pagebreak
\section{Acknowledgments}
No funding was received for conducting this study. The authors have no relevant financial or non-financial interests to disclose.

\section{Compliance with Ethical Standards}
This is a computational study on a publicly available benchmark for which no ethical approval was required.

{\small
\bibliographystyle{IEEEbib}
\bibliography{refs}

\begin{thebibliography}{10}

\bibitem{li2023sheetcopilot}
Hongxin Li, Jingran Su, Yuntao Chen, Qing Li, and Zhaoxiang Zhang,
\newblock ``{SheetCopilot}: Bringing software productivity to the next level
  through large language models,''
\newblock in {\em Proc. NeurIPS}, 2023.

\bibitem{wang2026sheetbrain}
Ziwei Wang, Jiayuan Su, Mengyu Zhou, Huaxing Zeng, Mengni Jia, Xiao Lv, Haoyu
  Dong, Xiaojun Ma, Shi Han, and Dongmei Zhang,
\newblock ``{SheetBrain}: A neuro-symbolic agent for accurate reasoning over
  complex and large spreadsheets,''
\newblock {\em Proc. AAAI Conf. Artif. Intell.}, vol. 40, 2026.

\bibitem{yang2025agentoccam}
Ke~Yang, Yao Liu, Sapana Chaudhary, Rasool Fakoor, Pratik Chaudhari, George
  Karypis, and Huzefa Rangwala,
\newblock ``{AgentOccam}: A simple yet strong baseline for {LLM}-based web
  agents,''
\newblock in {\em Proc. ICLR}, 2025.

\bibitem{wang2024codeact}
Xingyao Wang, Yangyi Chen, Lifan Yuan, Yizhe Zhang, Yunzhu Li, Hao Peng, and
  Heng Ji,
\newblock ``Executable code actions elicit better {LLM} agents,''
\newblock in {\em Proc. ICML}, 2024.

\bibitem{yang2024sweagent}
John Yang, Carlos~E. Jimenez, Alexander Wettig, Kilian Lieret, Shunyu Yao,
  Karthik Narasimhan, and Ofir Press,
\newblock ``{SWE-agent}: Agent-computer interfaces enable automated software
  engineering,''
\newblock in {\em Proc. NeurIPS}, 2024.

\bibitem{dietterich1998tests}
Thomas~G. Dietterich,
\newblock ``Approximate statistical tests for comparing supervised
  classification learning algorithms,''
\newblock {\em Neural Computation}, vol. 10, no. 7, 1998.

\bibitem{ma2024spreadsheetbench}
Zeyao Ma, Bohan Zhang, Jing Zhang, Jifan Yu, Xiaokang Zhang, Xiaohan Zhang,
  Sijia Luo, Xi~Wang, and Jie Tang,
\newblock ``{SpreadsheetBench}: Towards challenging real world spreadsheet
  manipulation,''
\newblock in {\em Proc. NeurIPS D\&B}, 2024.

\bibitem{zhu2026spreadsheetbench2}
Jian Zhu, Yuzheng Zhang, Zeyao Ma, Bohan Zhang, Armin Schoepf, Daniel Woloch,
  Peter~Yiliu Wang, Guangyu~Robert Yang, Samuel Jacob, Siddharth Nagisetty,
  Abhiram Chundru, Jean Lin, Spencer Mateega, and Jing Zhang,
\newblock ``{SpreadsheetBench 2}: Evaluating agents on end-to-end business
  spreadsheet workflows,''
\newblock {\em arXiv:2606.29955}, 2026.

\bibitem{schick2023toolformer}
Timo Schick, Jane Dwivedi-Yu, Roberto Dess{\`i}, Roberta Raileanu, Maria
  Lomeli, Eric Hambro, Luke Zettlemoyer, Nicola Cancedda, and Thomas Scialom,
\newblock ``Toolformer: Language models can teach themselves to use tools,''
\newblock in {\em Proc. NeurIPS}, 2023.

\bibitem{patil2024gorilla}
Shishir~G. Patil, Tianjun Zhang, Xin Wang, and Joseph~E. Gonzalez,
\newblock ``Gorilla: Large language model connected with massive {APIs},''
\newblock in {\em Proc. NeurIPS}, 2024.

\bibitem{shinn2023reflexion}
Noah Shinn, Federico Cassano, Ashwin Gopinath, Karthik Narasimhan, and Shunyu
  Yao,
\newblock ``Reflexion: Language agents with verbal reinforcement learning,''
\newblock in {\em Proc. NeurIPS}, 2023.

\bibitem{madaan2023selfrefine}
Aman Madaan, Niket Tandon, Prakhar Gupta, Skyler Hallinan, Luyu Gao, Sarah
  Wiegreffe, Uri Alon, Nouha Dziri, Shrimai Prabhumoye, Yiming Yang, et~al.,
\newblock ``Self-refine: Iterative refinement with self-feedback,''
\newblock in {\em Proc. NeurIPS}, 2023.

\bibitem{wang2024mobileagent}
Junyang Wang, Haiyang Xu, Jiabo Ye, Ming Yan, Weizhou Shen, Ji~Zhang, Fei
  Huang, and Jitao Sang,
\newblock ``{Mobile-Agent}: Autonomous multi-modal mobile device agent with
  visual perception,''
\newblock {\em arXiv:2401.16158}, 2024.

\bibitem{wang2024mobileagentv2}
Junyang Wang, Haiyang Xu, Haitao Jia, Xi~Zhang, Ming Yan, Weizhou Shen,
  Ji~Zhang, Fei Huang, and Jitao Sang,
\newblock ``{Mobile-Agent-v2}: Mobile device operation assistant with effective
  navigation via multi-agent collaboration,''
\newblock in {\em Proc. NeurIPS}, 2024.

\bibitem{huang2024selfcorrect}
Jie Huang, Xinyun Chen, Swaroop Mishra, Huaixiu~Steven Zheng, Adams~Wei Yu,
  Xinying Song, and Denny Zhou,
\newblock ``Large language models cannot self-correct reasoning yet,''
\newblock in {\em Proc. ICLR}, 2024.

\bibitem{tyen2024errorlocation}
Gladys Tyen, Hassan Mansoor, Victor {C\u{a}rbune}, Peter Chen, and Tony Mak,
\newblock ``{LLM}s cannot find reasoning errors, but can correct them given the
  error location,''
\newblock in {\em Findings of ACL}, 2024.

\bibitem{zhou2023leasttomost}
Denny Zhou, Nathanael Sch{\"a}rli, Le~Hou, Jason Wei, Nathan Scales, Xuezhi
  Wang, Dale Schuurmans, Claire Cui, Olivier Bousquet, Quoc Le, and Ed~Chi,
\newblock ``Least-to-most prompting enables complex reasoning in large language
  models,''
\newblock in {\em Proc. ICLR}, 2023.

\bibitem{khot2023decomposed}
Tushar Khot, Harsh Trivedi, Matthew Finlayson, Yao Fu, Kyle Richardson, Peter
  Clark, and Ashish Sabharwal,
\newblock ``Decomposed prompting: A modular approach for solving complex
  tasks,''
\newblock in {\em Proc. ICLR}, 2023.

\bibitem{prasad2024adapt}
Archiki Prasad, Alexander Koller, Mareike Hartmann, Peter Clark, Ashish
  Sabharwal, Mohit Bansal, and Tushar Khot,
\newblock ``{ADaPT}: As-needed decomposition and planning with language
  models,''
\newblock in {\em Findings of NAACL}, 2024.

\bibitem{chen2024morecalls}
Lingjiao Chen, Jared~Quincy Davis, Boris Hanin, Peter Bailis, Ion Stoica, Matei
  Zaharia, and James Zou,
\newblock ``Are more {LLM} calls all you need? towards the scaling properties
  of compound {AI} systems,''
\newblock in {\em Proc. NeurIPS}, 2024.

\bibitem{cemri2025masfail}
Mert Cemri, Melissa~Z. Pan, Shuyi Yang, Lakshya~A. Agrawal, Bhavya Chopra,
  Rishabh Tiwari, Kurt Keutzer, Aditya Parameswaran, Dan Klein, Kannan
  Ramchandran, Matei Zaharia, Joseph~E. Gonzalez, and Ion Stoica,
\newblock ``Why do multi-agent {LLM} systems fail?,''
\newblock in {\em Proc. NeurIPS D\&B}, 2025.

\bibitem{card2020littlepower}
Dallas Card, Peter Henderson, Urvashi Khandelwal, Robin Jia, Kyle Mahowald, and
  Dan Jurafsky,
\newblock ``With little power comes great responsibility,''
\newblock in {\em Proc. EMNLP}, 2020.

\bibitem{dong2024maxmind}
Yuchen Dong, Xiaoxiang Fang, Yuchen Hu, Renshuang Jiang, and Zhe Jiang,
\newblock ``{MaxMind}: A memory loop network to enhance software productivity
  based on large language models,''
\newblock {\em arXiv:2408.03841}, 2024.

\bibitem{wu2024oscopilot}
Zhiyong Wu, Chengcheng Han, Zichen Ding, Zhenmin Weng, Zhoumianze Liu, Shunyu
  Yao, Tao Yu, and Lingpeng Kong,
\newblock ``{OS-Copilot}: Towards generalist computer agents with
  self-improvement,''
\newblock {\em arXiv:2402.07456}, 2024.

\bibitem{liu2024agentbench}
Xiao Liu, Hao Yu, Hanchen Zhang, Yifan Xu, Xuanyu Lei, Hanyu Lai, Yu~Gu,
  Hangliang Ding, Kaiwen Men, Kejuan Yang, et~al.,
\newblock ``{AgentBench}: Evaluating {LLM}s as agents,''
\newblock in {\em Proc. ICLR}, 2024.

\end{thebibliography}
}

\end{document}